\documentclass{bioont}

\usepackage{natbib} \usepackage{url} \usepackage{listings}
\usepackage{xspace}

\lstMakeShortInline[]|

\begin{document}

\title{The Karyotype Ontology: a computational representation for human
  cytogenetic patterns}

\author{Jennifer D. Warrender and Phillip Lord\footnote{To whom
    correspondence should be addressed: phillip.lord@newcastle.ac.uk}}

\address{School of Computing Science, Newcastle University,
Newcastle-upon-Tyne, UK}

\date{March 2013}

\maketitle

\begin{abstract}
  The karyotype ontology describes the human chromosome complement as
  determined cytogenetically, and is designed as an initial step toward the
  goal of replacing the current system which is based on semantically
  meaningful strings. This ontology uses a novel, semi-programmatic
  methodology based around the \textit{tawny} library to construct many
  classes rapidly. Here, we describe our use case, methodology and the
  event-based approach that we use to represent karyotypes.

  The ontology is available at
  \url{http://www.purl.org/ontolink/karyotype/}. The clojure
  code is available at \url{http://code.google.com/p/karyotype-clj/}.
\end{abstract}

\section{Introduction}
\label{sec:introduction}

A karyotype describes the chromosome complement of the individual. It can be
easily assayed cytogenetically and, therefore, has been widely used as a
mechanism for understanding the underlying genetic complement of cells and
organisms. It remains of vital diagnostic importance, as well as a key tool
for a large research community. Human karyotypes are normally represented
using a string, as defined by the International System for human Cytogenetic
Nomenclature 2009 (ISCN2009)~\citep{shaffer09} --- here we call these
\textit{ISCN strings}. Unlike similar string-based representations such as
InCHI~\citep{mcnaught07}, ISCN strings lack a formal interpretation, and do
not have good computational properties. For example, they cannot be
represented in ASCII as they include meaningful underlining. Even the ISCN
specification has no electronic representation and is not searchable.

In this paper, we describe our work in developing an ontological
representation for karyotypes. Currently, karyotypes have only been
represented as experimental entities, or results of medical
procedures.  The purpose of our ontology is to provide a strong
computational and formal interpretation for a karyotype. This will
enable semantic (and syntactic) checking of karyotypic information at
the point of generation; it will allow the development of a knowledge
base of karyotypes which is open to rich querying, and finally a
web-capable interchange format as many different groups around the
world generate this information.

\section{What is an ISCN string}
\label{sec:what-an-iscn}

ISCN defines a string format, initially designed for writing and printing,
which provides a representation of the chromosome complement of a human, as
determined cytogenetically by the banding patterns which are revealed after
staining and fixation of metaphase cells. The ISCN specification has a long
history. Initially, it was developed to address the need for an explicit
nomenclature ``to enable communication between workers in the field''. The
early versions date from around 1960, when the emphasis was on human-to-human
communication, and for small numbers of karyotypes. 

ISCN strings represents a number of key concepts: 
\begin{itemize}
\item The autosomal chromosomes are represented by a number \texttt{1}
  to \texttt{22}.
\item The sex chromosomes are represented by \texttt{X} or \texttt{Y}.
\item Chromosomal structural components are represented: The long and
  short arm are represented by \texttt{q} and \texttt{p};
  centromeres are represented by \texttt{cen} or more specifically,
  \texttt{p10} for the part of the centromere facing the short arm or
  \texttt{q10} for part facing the long arm; and
  telomeres are represented by \texttt{ter}.
\item Bands at different resolutions are represented numerically such
  as \texttt{1p11.1}.
\item Changes from the base karyotypes are represented: \texttt{del}
  represents a deletion, \texttt{add} represents additional material. 
\end{itemize}

In addition to these concepts, there are many more that can be represented in
an extended karyotype: these include chromosomal groups, mosaicism, ploidy
level and so forth. The full specification, describes many parts of human
cytogenetics, including both the biology and the experimental techniques used.
As would be expected for a specification with a long history, not all parts
are regularly used.

\textit{Banding Patterns} used to describe chromosome locations are defined
cytogenetically by the appearance of the chromosome, during a part of
division, following staining with a dye; this staining process is normally
lethal to the cell. The original banding pattern described in the Paris
Conference 1971 report, represented the results of three whole chromosome
banding techniques: Quinacrine- (Q-), Giemsa- (G-), and Reverse-banding (R-).

The main components of stained chromosomes are: 
\begin{itemize}
\item A \texttt{band} is a part of the chromosome that is
  distinguishable from its adjacent segments, appearing darker or
  lighter. Bands proximal to the centromere are labelled as 1, then 2
  and so on.
\item A \texttt{region} is an area of a chromosome lying between two
  landmarks. Regions adjacent to the centromere are labelled as 1 in
  each arm, then 2 and so on.
\item A \texttt{landmark} is a consistent and distinct morphological
  feature of a chromosome. They are used as a delimiters for regions.
\end{itemize}

Bands are represented numerically such that bands are numbered from the
centromere outward. The band name is a combination of: the chromosome number,
the arm symbol, the region number, and the band number within that region. For
example, the band \texttt{1q42} is found on the long arm of chromosome 1 and
is the second band, proximal to the centromere in region 4. Broadly speaking
there is a correlation between the cytogenetic bands and the underlying DNA
sequence of the chromosome; however, the very different scales (\texttt{1q42}
is 12.4Mb long) of these two measurements means that this relationships is
approximate.

Cytogenetic banding also comes at several resolutions: high-resolution
banding involves the staining of chromosomes during prophase, prometaphase, or
interphase when the chromosomes are less condensed and spread over a large
area; low-resolution uses the highly-condensed metaphase chromosome. The level
of resolution is determined by the number of bands seen in a haploid set and
ranges approximately from 300 to 850. High-resolution banding techniques
result in existing bands being subdivided into \texttt{sub-bands}. Whenever a
band is subdivided, a decimal place is placed after the original band number,
and the sub-band number is appended to the band name with proximal sub-bands
bring labelled 1, then 2 and so on. For example, the sub-bands of band
\texttt{1q42} will be: \texttt{1q42.1}, \texttt{1q42.2}, and \texttt{1q42.3},
such that sub-band \texttt{1q42.1} is more proximal to the centromere. However
when a sub-band is subdivided, then no additional decimal is added. For
example, the sub-bands of sub-band \texttt{1q42.1} are: \texttt{1q42.11},
\texttt{1q42.12}, and \texttt{1q42.13}.

Typical queries that we might wish to make of a collection of karyotypes include:
\begin{itemize}
\item Which karyotypes have abnormalities in a given chromosome?
\item Which karyotypes increase the copy number of a given band?
\item Which karyotypes affects a given band in any way?
\end{itemize}
Currently, these are hard to answer computationally because of the complexity
of the ISCN strings, as well as intrinsic complexity of the biology. Our
karyotype ontology aims to address the former, and contain the latter.

\section{Our Methodology}
\label{sec:our-methodology}

For the karyotype ontology, we have a very specific requirement which is to
enable machine interpretation of the knowledge that is currently represented
in ISCN strings. One of the implications of this is that our knowledge capture
is extremely contained; essentially all the knowledge we require is present in
the ISCN2009 specification\footnote{During the course of the work described
  here, ISCN2013 was released (in 2012!). We have not updated for this version
  yet.}. Our task is to formalise and represent this.

Our initial experiments with a realist ontology showed a number of
difficulties; the distinction between a chromosome (as a piece of DNA and
protein), the experimental artefact (following staining) and the visualisation
of the experimental artefact are all different ``portions of reality''; for
instance, a chromosome in a live cell cannot meaningfully be said to have
bands. These distinctions can be represented ontologically, however the result
is a ontology with many duplicated hierarchies: chromosome 1, stained
chromosome 1, and the visualisation of chromosome 1.

As these distinctions are not required for our application, we have instead
followed a pragmatic approach~\citep{reality_in_biology_2010}. We have developed
a lightweight ontology with specific computational goals. Our desire for
computational support and inferencing, as well as a web-capable interchange
format, has lead us to adopt OWL2 as our representation format.

While avoiding a realist approach has reduced some duplication, karyotype
ontology still requires a considerable number of highly similar concepts,
which is intrinsic to the problem domain. Trivially, for instance, the human
karyotype requires 24 individual chromosome concepts, with similar logical and
textual definitions. In turn, each chromosome has a complex band pattern (over
850 bands in total), including band intervals for use at different
resolutions. Developing this type of ontology would be complex, time consuming
and difficult to maintain using conventional tools. Therefore, we developed
the \textit{tawny} library, which allows fully programmatic development of OWL
ontologies~\citep{greycite8873}. This library allows expansion of arbitrary
patterns; this is similar to the capabilities of
OPPL\citep{aranguren_Stevens_Antezana_2009}, populous~\citep{jupp10} or safe
macros~\citep{mungall10}. However, it additionally provides us with the
ability to define unit tests to provide computationally checkable expressions
of the requirements for inferencing, a semi-literate programming environment,
and the ability to make arbitrary syntactic extensions. The syntax is modelled
after and highly similar to Manchester syntax, therefore it is presented here
without further explanation; a fuller description is provided in the tawny
documentation\footnote{\url{https://github.com/phillord/tawny-owl}}.

\begin{figure}[!htpb]
\centering 
\includegraphics[scale=0.45]{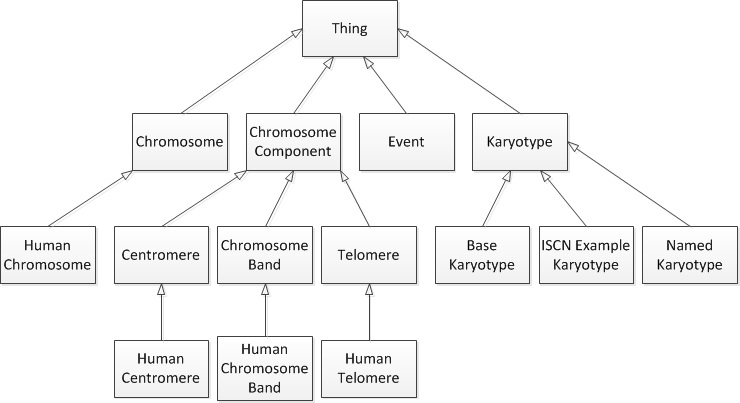}
\caption{The abstract structure of the karyotype ontology.}
\label{fig:structure}
\end{figure}

The basic structure of our ontology is shown in Figure~\ref{fig:structure}. 

While ISCN2009 contains most of the knowledge we need, it is informally
represented; it sometimes lacks the clarity and contains contradictions;
trivially, for instance, Group G is described as having satellites (page 7),
while Chromosome Y (a member of Group G) is described as having no satellites.
A similar confusion lies over the instance/type distinction. So, while for an
individual organism or cell line all (normal) Chromosome 21s have (visible)
satellites, an individual chromosome, in an individual cell may
not\footnote{As well as variation at a genetic level in even a clonal
  population, the cytogenetic definition of satellite is a staining region. So
  even for two genetically identical chromosomes, one may show a satellite,
  and another may not.}. Similarly small supernumerary marker chromosomes
(sSMCs) cannot be fully represented in ISCN2009 ~\citep{liehr09}. This means
that an ontological representation of the ISCN2009 specification cannot
therefore be a completely faithful representation.

\section{Representing Abnormalities}
\label{sec:repr-abnorm}

Our initial experiments attempt to represent karyotypes as a rich
partonomy, using the concepts described in the previous section. For
example, the normal male karyotype \texttt{46,XY} would be described
as having 46 chromosomes of the appropriate types. One problem with
this approach is that the definition of any karyotype is relatively
large; while tawny enables the generation of this form of concept,
it cannot reduce the complexity for reasoners; so this form of
ontology is not likely to scale well. Further, a simple partonomy is
not a rich enough representation, as it cannot represent simple
inversions which contain all the same parts, but not necessarily in
the right order. While it is possible to represent order in
OWL~\citep{drummond06}, this would add more complexity and scaling issues.

Further, there are some strong edge cases that are not complex, but impossible
to represent as a partonomy; for example, the karyotype \texttt{45,X,-Y}
describes the chromosomes of a cell line, isolated from a normal male, which
has lost its Y chromosome. This is a different karyotype but is partonomically
indistinguishable from \texttt{45,X}, a phenotypically female individual with
Turners syndrome. Therefore we represented karyotypes using events: a
karyotype is described as a set of changes from a \textit{base} or normal
karyotype. For example, \texttt{45,X,-Y} is described as a \texttt{46,XY}
male, with a deletion event of \texttt{Y}, while \texttt{45,X} female is
described as a \texttt{46,XN}, with a deletion event of a sex chromosome (see
Listing\ref{lst:k45_X}): in this case N represents an unknown autosome. OWL
can represent partial knowledge straight-forwardly.

\begin{lstlisting}[caption={A basic class definition},label=lst:k45_X]
(defclass k45_X
  :label "The 45,X karyotype"
  :subclass ISCNExampleKaryotype
  (owlsome derivedFrom k46_XN)
  (deletion 1 HumanSexChromosome))
\end{lstlisting}

In total there are 13 events that represent the following key concepts:
\begin{itemize}
\item\texttt{Addition}: Chromosome gain or band addition.
\item\texttt{Deletion}: Chromosome loss or band deletion, including terminal
  deletion with break (:) and interstitial deletion with breakage and reunion
  (::).
\item\texttt{Duplication}: Band duplication. Specialised with
  \\\texttt{DirectDuplication} or \texttt{InverseDuplication}.
\item\texttt{Fisson}: Centric fission.
\item\texttt{Insertion}: Band insertion between chromosomes. Specialised with 
  \texttt{DirectInsertion} or \texttt{InverseInsertion}.
\item\texttt{Inversion}: Band inversion, both paracentric and
  pericentric.
\item\texttt{Quadruplication}: Band quadruplication.
\item\texttt{Translocation}: Band translocation between chromosomes.
\item\texttt{Triplication}: Band triplication.
\end{itemize}

These concepts are supported by a number of properties created for the
purpose, such as \texttt{isBandOf}.

The simplest representation in our ontology shares some limitations with the
ISCN ``short system'' -- a usable subset of ISCN, which is more generally
used. For example, with both triplication or quadruplication events, we do not
represent the orientation of all the repeats. It would be possible to
differentiate these as two direct duplications, or one direct and one inverted
duplication (for triplication); however, again it is useful to represent
partial information; for many existing ISCN strings which use the short
system, this knowledge is not available.

\section{Defining Sex}
\label{sec:sex}

One interesting outcome of both our representation and normal custom within
cytology is that the definition of the sex of a karyotype is quite different
from what might be expected. Intuitively, male would be defined as a karyotype
with a Y chromosome\footnote{We ignore Y chromosome translocations for
  simplicity}, while female would be defined as a karyotype without. However,
this intuitive definition is not correct. For example, the previously
described \texttt{45X,-Y} has no chromosome Y and yet would generally be
considered to be a male karyotype, since the organism from which the cell line
originated was male. Our definitions of male and female, therefore, consider
the ``history'' of the karyotype. Female is defined as derived from
\texttt{46,XX} (Listing\ref{lst:female}), Male from \texttt{46,XY}
(Listing\ref{lst:male}). This definition also copes with Turner's syndrome
which is not defined as either male or female, nor describe sex for haploid
karyotypes: these can contain a Y chromosome (or not), but sex is not
meaningful for these karyotypes. Karyotypes which are definitional for
syndromes such as Turner's or being male, are categorised under
\texttt{NamedKaryotype}. 

\begin{lstlisting}[caption={Definition of Female Karyotype},label=lst:female]
(defclass FemaleKaryotype
  :equivalent
  (owlor k46_XX
         (owlsome derivedFrom k46_XX)))
\end{lstlisting}

\begin{lstlisting}[caption={Definition of Male Karyotype},label=lst:male]
(defclass MaleKaryotype
  :equivalent
  (owlor k46_XY
         (owlsome derivedFrom k46_XY)))
\end{lstlisting}

While this is ontologically correct, it does remove some inferencing power
that might reasonably be expected; the karyotype \texttt{45,X} is effectively
stated to be female, as it is not formally possible to determine the sex from
the components of a karyotype; for future work, we may be able to address
this, by describing the origin of the karyotype (\texttt{45,X,-Y} is only
valid as the karyotype of a cell line).

\section{Assessment}
\label{sec:assessment}

As well as providing a specification, we are fortunate that ISCN2009 provides
many examples; we are using these examples as an initial evaluation for our
ontology, to determine whether the ontology is expressive enough to represent
these exemplar karyotypes. 

We wished this to be related to the ISCN string as, in most cases, there is no
other more humanly readable name. In order to represent a karyotype in
tawny a name is needed which is ``safe'' both as a URL and in Clojure, the
language used to implement tawny, and, pragmatically, in Manchester syntax
also\footnote{It is possible to dissociate these two with tawny, but that
  did not seem useful in this case}.

\begin{itemize}
\item All karyotypes start with a ``k'' --- Clojure symbols cannot start with
  numbers
\item Replaced ; character with \_ --- comment in Clojure
\item Replaced ( and ) characters with ! --- list delimiter in Clojure
\item Replaced , character with \_ --- separator in Manchester syntax
\end{itemize}

Currently, we have represented 71 karyotypes in our karyotype
ontology. During this process, we have also discovered two
difficulties with the existing ISCN2009 specification; in both cases a
simple and intuitive correction is possible. These are:

\begin{itemize}
\item The lack of a band \texttt{Xq12} in figures showing chromosome
  bands (page 31); the figures are the only list of all chromosome
  bands in ISCN2009.
\item The absence of a band \texttt{Yq11.2} in the 300 band resolution
  (\texttt{11.21}, \texttt{11.22}, \texttt{11.23} do exist on page 31)
  while this band is used in several exemplars (for example on page
  78). This band does exist in ISCN2005 -- the previous specification.
\end{itemize}

Taken together, these 71 karyotypes use all of the distinctions necessary to
answer questions given in Section~\ref{sec:what-an-iscn}.

\section{Discussion}
\label{sec:discussion}

The development of a karyotype ontology is potentially valuable for
cytogenetics, as the current ISCN specification is not computationally
amenable, reducing the value of collections of karyotypes as they are hard to
query, check and maintain. The work described here presents an initial step
towards this goal. The process of producing this ontology is already of use; we
have discovered some errors or inconsistencies within ISCN which prevent its
direct interpretation computationally; we expect to find more as we continue.

We have found the use of an ontology to be an appropriate mechanism; the
knowledge that needs to be represented is complex, and overlapping.
Cytogenetic data also requires the representation of partial knowledge, such
as locations that are only known to a given resolution. The open world
assumption of OWL copes well with this situation. Cytogenetics databases are
also relatively small (100,000's rather than millions or billions), sizes to
which OWL should scale.

Existing tools for ontology development are, however, rather limited
in their support for building this form of ontology; it is to address
this need that we have developed tawny. This has proven to be
highly useful; for example, the current karyotype ontology consists of
1466 classes, of which 1293 are used to represent the chromosomes and
their bands at different resolutions.  All of these classes have been
generated from simpler data structures in Clojure. Additionally, the
arbitrary expressiveness of tawny has allowed us to add syntax
specific for the karyotype; many definitions in our ontology follow
patterns. Using tawny these can be encoded as Clojure functions,
such as that shown in and used in Listing\ref{lst:inverse}.

\begin{lstlisting}[caption={A function used to define inverse events},label=lst:inverse]
(defn inversion
  [n band1 band2]
    (exactly n hasEvent 
        (owland Inversion 
            (owlsome hasBreakPoint 
                     band1 band2))))

(e/inversion 1 h/Hu2Bandp21 h/Hu2Bandq31)
\end{lstlisting}

In addition to the convenience, this also aids significantly in
maintainability, as it is possible to change definitions for all classes that
use this function. In time, we expect to extend this work to present an
end-user syntax and parser, probably built directly using Clojure. Through the
use of tawny we aim to build an end-to-end solution for the computational
encoding of karyotypes.

\section*{Acknowledgements}

This work was supported by Newcastle University.

\bibliographystyle{natbib}

\bibliography{mybib,phil_lord_refs}

\end{document}